\documentclass{aptpub}

\authornames{Wang and Resnick} 
\shorttitle{Reciprocity in PA} 

\usepackage{color,graphicx,verbatim,amsmath,verbatim, amsfonts, url,cases,tikz,epsfig}

\newcommand{\PP}{\mathbb{P}}
\newcommand{\EE}{\mathbb{E}}
\newcommand{\dd}{\mathrm{d}}
\newcommand{\ind}{\boldsymbol{1}}

\numberwithin{equation}{section}
\numberwithin{figure}{section}
\numberwithin{theorem}{section}
\numberwithin{corollary}{section}
\numberwithin{lemma}{section}
\usepackage[numbers,sort&compress]{natbib}

\newcommand{\deltain}{\delta_\text{in}}
\newcommand{\deltaout}{\delta_\text{out}}
\newcommand{\Din}{D^\text{in}}
\newcommand{\Dout}{D^\text{out}}

\newcommand{\convas}{\stackrel{\text{a.s.}}{\longrightarrow}}

\newtheorem{Remark}{Remark}[section]
\allowdisplaybreaks

\begin{document}

\title{Measuring Reciprocity in a Directed Preferential Attachment Network} 

\authorone[Texas A\&M University]{Tiandong Wang} 
\authortwo[Cornell University]{Sidney I. Resnick}

\addressone{Department of 
	Statistics, Texas 
	A\&M University, College 
		Station, TX 77843, U.S.} 
\emailone{twang@stat.tamu.edu} 
\addresstwo{School of Operations Research and Information Engineering,
	 Cornell University, Ithaca, NY 14853, U.S.}
	 \emailtwo{sir1@cornell.edu}

\begin{abstract}
Empirical studies \cite{jiangetal:2015, social:2007} show that online social networks have not only in- and out-degree
distributions with Pareto-like tails, but also a high proportion of reciprocal edges.
A classical directed preferential attachment (PA) model generates in- and out-degree distribution 
with power-law tails, but theoretical properties of the reciprocity feature in this model have not yet been studied. We derive the asymptotic results on the number of reciprocal edges between two fixed nodes, as well as the proportion of reciprocal edges in the entire PA network. We see that with certain choices of parameters, the proportion of reciprocal edges in a directed PA network is close to 0, which differs from the empirical observation. 
This points out one potential problem of fitting a classical PA model to a given network dataset with high reciprocity, and indicates alternative models need to be considered. 
\end{abstract}

\keywords{Reciprocity, preferential attachment, in- and out-degrees.}

\ams{60F15; 60G46}{05C82} 



\section{Introduction}
In social network analysis,  
reciprocated edges characterize the communication between two users.
For instance, on Facebook, one user leaves messages on another user's wall page, and the response from the target user then creates a reciprocal edge. 
\emph{Reciprocity}, which is classically defined as the proportion of reciprocated edges (cf. \cite{newman:etal:2001, wasserman:faust:1994}), is
one important network metric to measure interactions among individual users.
The directed network constructed from Facebook wall posts \cite{facebook:2009} is one example
of social networks with a large proportion of reciprocal edges. 
The study on eight different types of networks in \cite{jiangetal:2015} shows online social networks (e.g. \cite{facebook:2009, flickr:2009, java:etal:2007, googleplus:2012,social:2007}) tend to have a higher proportion of reciprocal edges, compared to other types of networks such as biological networks, communication networks, software call graphs and P2P networks. 

Another widely observed feature of directed social networks is the scale-free property, where both in- and out-degree distributions have Pareto-like tails. 
The directed preferential attachment (PA) network model is appealing (cf. \cite{bollobas:borgs:chayes:riordan:2003, krapivsky:redner:2001}), since theoretically the directed PA mechanism
generates a network with the scale-free property because nodes
with large degrees are likely to attract more edges than those with small degrees 
(cf. \cite{resnick:samorodnitsky:towsley:davis:willis:wan:2016, resnick:samorodnitsky:2015, wan:wang:davis:resnick:2017,wang:resnick:2019}). 
However, asymptotic behavior of the proportion of reciprocal edges in a directed PA model has not yet been explored in the literature. In this paper, we derive asymptotic results about: (1) the number of reciprocal edges between two fixed nodes; (2) the proportion of reciprocal edges in a directed PA network, provided that the network has a large number of edges.

Our theoretical results suggest that for certain choices of model parameters, especially when the proportion of edges added between two exisiting nodes is small, the proportion of reciprocal edges in the entire graph is close to 0, even though the total number of reciprocal edges between two fixed nodes may be of order $O(n^a)$, $a\in (0,1)$. Such behavior flags potential problems for fitting a directed PA model in practice. When fitting a directed PA model to a real network with high reciprocity using existing methods developed in \cite{wan:wang:davis:resnick:2017, wan:wang:davis:resnick:2017b}, 
there is no guarantee that the calibrated model will also have a high proportion of reciprocal edges.
Such discrepancy indicates a poor fit of the PA model, since the fitted model fails to capture the important feature of high reciprocity. In these cases, variants of the directed PA model need to be considered. For instance, \cite{chengetal:2011} provides several different ways to predict the reciprocal edges between two given nodes, and one may incorporate those features to construct a refined network model that is both scale-free and of high reciprocity.

The rest of the paper is organized as follows.
We provide an overview of the evolution in a directed PA model in Section~\ref{sec:model}.
Then we collect useful preliminary results on the growth of in- and out-degrees for a fixed node in Section~\ref{sec:prelim} with detailed proofs given in Appendix~\ref{sec:pf}. 
In Section~\ref{sec:recip}, we derive the limiting behavior of: (1) the number of reciprocal edges between two fixed nodes; (2) the first time when a reciprocal pair of edges is formed between two fixed nodes; (3) the proportion of reciprocal edges in a given directed PA model.
Discussion of the theoretical results and future research directions are then summarized in Section~\ref{sec:discuss}.

\subsection{Model Setup}\label{sec:model}
We now outline the basic setup of the directed PA model.
Initialize the model with graph $G(0)$, which consists of one node (labeled as Node 1) and a self-loop.
Let $G(n)$ denote the graph  after $n$ steps and
 $V(n)$ be the set of nodes in $G(n)$ with $V(0) = \{1\}$ and $|V(0)| = {1}$.
Denote the set of directed edges in $G(n)$ by $E(n)$ such that
an ordered pair $(i,j)\in E(n)$, $i,j\in V(n)$, represents a directed edge $i\mapsto j$. 
When $n=0$, we have $E(0) = \{(1,1)\}$.

Set $\bigl(\Din_v(n), \Dout_v(n)\bigr)$ to
be the in- and out-degrees of node $v$ in $G(n)$. 
We use the convention that $\Din_i(n) = 0$ if $i\notin V(n)$,
and similarly, $\Dout_j(n) = 0$ if $j\notin V(n)$.
From $G(n)$ to $G(n+1)$, one of the three scenarios happens:
\tikzset{
    >=stealth,
    punkt/.style={
           rectangle,
           rounded corners,
           draw=black, very thick,
           text width=6.5em,
           minimum height=2em,
           text centered},
    pil/.style={
           ->,
           thick,
           shorten <=2pt,
           shorten >=2pt,}
}
\bigskip
  \begin{center}
  \begin{tikzpicture}
    \begin{scope}[xshift=0cm,yshift=1cm]
      \node[draw,circle,fill=white] (s1) at (2,0) {$v$};
      \node[draw,circle,fill=gray!30!white] (s2) at (.5,-1.5) {$w$};
      \draw[->] (s1.south west)--(s2.north east){};
      \draw[dashed] (0,-2.2) circle [x radius=2cm, y radius=15mm];
    \end{scope}
    
     \begin{scope}[xshift=5cm,yshift=1cm]
      \node[draw,circle,fill=gray!30!white] (s1) at (.5,-1.5) {$v$};
      \node[draw,circle,fill=gray!30!white] (s2) at (-.5,-2.5) {$w$};
      \draw[->] (s1.south west)--(s2.north east){};
      \draw[dashed] (0,-2.2) circle [x radius=2cm, y radius=15mm];
    \end{scope}
    
     \begin{scope}[xshift=10cm,yshift=1cm]
      \node[draw,circle,fill=white] (s1) at (2,0) {$v$};
      \node[draw,circle,fill=gray!30!white] (s2) at (.5,-1.5) {$w$};
      \draw[->] (s2.north east)--(s1.south west){};
      \draw[dashed] (0,-2.2) circle [x radius=2cm, y radius=15mm];
    \end{scope}
    
     \node at (0,-3.5) {$\alpha$-scheme};
     \node at (5,-3.5) {$\beta$-scheme};
     \node at (10,-3.5) {$\gamma$-scheme};
  \end{tikzpicture}    
  \end{center}  
\bigskip
\begin{enumerate}
\item With probability $\alpha$, we add a new edge $(w,v)$, where $v\in V(n)$, and
 $w\notin V(n)$ is a new node. The existing node $v$ is chosen with probability 
\[
\frac{\Din_v(n)+\deltain}{\sum_{v\in V(n)} (\Din_v(n)+\deltain)}
= \frac{\Din_v(n)+\deltain}{n+1+\deltain |V(n)|}. 
\]
\item With probability $\beta$, a new edge $(w,v)$ is added between two existing nodes $w,v\in V(n)$, where the starting and the ending nodes $w, v$ are chosen independently with probability
\begin{align*}
&\frac{\Din_v(n)+\deltain}{\sum_{v\in V(n)} (\Din_v(n)+\deltain)}\frac{\Dout_w(n)+\deltaout}{\sum_{w\in V(n)} (\Dout_w(n)+\deltaout)} \\
&\qquad= 
\frac{\Din_v(n)+\deltain}{n+1+\deltain |V(n)|}\frac{\Dout_w(n)+\deltaout}{n+1+\deltaout |V(n)|}.
\end{align*}
For brevity of notation, we set 
\begin{equation}\label{eq:def_Aij}
A_{vw}(n) := \frac{\Din_v(n)+\deltain}{n+1+\deltain |V(n)|}\frac{\Dout_w(n)+\deltaout}{n+1+\deltaout|V(n)|},
\end{equation}
then the attachment probability in the $\beta$-scheme is $\beta A_{vw}(n)$.
\item With probability $\gamma$, we add a new edge $(v,w)$, where $v\in V(n)$, and
$w\notin V(n)$ is a new node.
 The existing node $v$ is chosen with probability 
\[
\frac{\Dout_v(n)+\deltaout}{\sum_{v\in V(n)} (\Dout_v(n)+\deltaout)}
=\frac{\Dout_v(n)+\deltaout}{n+1+\deltaout |V(n)|}. 
\]
\end{enumerate}
We assume $\alpha+\beta+\gamma=1$, $\beta\in [0,1)$, and $\deltain,\deltaout>0$.
Due to the $\alpha$- and $\gamma$-schemes, $|V(n)|-1$ follows a binomial
distribution with size $n$ and success probability $\alpha+\gamma=1-\beta$,
so that $|V(n)|\convas\infty$ as $n\to\infty$.
For $v\ge 1$, we define $S_v$ to be the time when node $v$ is created, i.e.
\begin{equation}\label{eq:def_Sv}
S_v := \inf\{n\ge 0: |V(n)| = v\}.
\end{equation}
Since we use the convention that $\Din_v(n) = 0$ and $\Dout_v(n) = 0$ if $S_v>n$, 
we have by \eqref{eq:def_Aij} that
$A_{vw}(n)\in [0,1]$, for all $n$.

\section{Preliminaries on $\Din_v(n)$, $\Dout_v(n)$}\label{sec:prelim}
Fix $i,j \in V(n)$, $n\ge 1$, and
write 
\begin{equation}\label{eq:c1c2}
c_1 = \frac{\alpha+\beta}{1+\deltain(1-\beta)}, \qquad c_2 = \frac{\beta+\gamma}{1+\deltaout(1-\beta)}.
\end{equation}
We see from \eqref{eq:c1c2} that
$0<c_1<\alpha+\beta $, $0<c_2<\beta+\gamma$, and $0<c_1+c_2 < 1+\beta$.
In fact, we can re-parametrize the PA model using $(\alpha, \beta,\gamma, c_1, c_2)$,
and later in Lemma~\ref{lemma:moment}, we show that
 $c_1, c_2$ control the growth rates of in- and out-degrees for a fixed node $v$, respectively.

Let $\mathcal{F}_n$ denote the $\sigma$-field generated by observing the network evolution up to 
the creation of $n$-th new edge. Suppose $\tau$ is a stopping time with respect to the filtration $(\mathcal{F}_n)_{n\ge 0}$, then 
\[
\mathcal{F}_{\tau} = \{F: F\cap\{\tau=n\}\in \mathcal{F}_n\}.
\]
By \eqref{eq:def_Sv}, we see that $S_v$ is a stopping time with respect to $(\mathcal{F}_n)_{n\ge 0}$. 
For $n\ge k\ge 0$, we have
\[
\{S_v+k=n\} = \{S_v=n-k\}\in \mathcal{F}_{n-k}\subset \mathcal{F}_n,
\]
so $S_v+k$, $k\ge 0$, is a stopping time with respect to $(\mathcal{F}_n)_{n\ge 0}$.
Note that for $v>n-k$, $\{S_v=n-k\}=\emptyset\in \mathcal{F}_{n-k}\subset \mathcal{F}_n$.
We now collect useful lemmas for later analyses, and their proofs are collected in Section~\ref{sec:pf}.

\begin{lemma}\label{lem:fil}
For some integer $p\ge 1$, $k\ge 0$, and $v\ge 1$, we have:
\begin{enumerate}
\item[(i)] For $p=1$, 
\begin{align}
\EE^{\mathcal{F}_{S_v+k}}&\left(\Din_v(S_v+k+1)+\deltain\right)\nonumber\\
&=(\Din_v(S_v+k)+\deltain)\left(1+\frac{\alpha+\beta}{S_v+k+1+\deltain|V(S_v+k)|}\right).
\label{eq:deg_1}
\end{align}
\item[(ii)] For an integer $p\ge 2$,
\begin{align}
&\EE^{\mathcal{F}_{S_v+k}}\left(\left(\Din_v(S_v+k+1)+\deltain\right)^p\right)\nonumber\\
&=
(\Din_v(S_v+k)+\deltain)^p\left(1+p\frac{\alpha+\beta}{S_v+k+1+\deltain|V(S_v+k)|}\right) \nonumber\\
&\qquad+ \frac{\alpha+\beta}{S_v+k+1+\deltain|V(S_v+k)|}\sum_{r=2}^p{p \choose r}(\Din_v(S_v+k)+\deltain)^{p-r+1}.
\label{eq:deg_p}
\end{align}
\end{enumerate}
\end{lemma}
The proof of Lemma~\ref{lem:fil} is given in Appendix~\ref{subsec:pf2.1}.

Next, we study properties of $\EE\left[(\Din_v(S_v+k))^p\right]$ and $\EE\left[(\Dout_v(S_v+k))^p\right]$, $k\ge 1$, which are useful preliminary results for deriving theorems in Section~\ref{sec:recip}.
\begin{lemma}\label{lemma:moment}
For $v\in V(n)$ and $p\ge 1$, we have
\begin{align*}
\sup_{k\ge 1}&\frac{\EE\left[(\Din_v(k))^p\right]}{k^{c_1p}}<\infty,\qquad
\sup_{k\ge 1}\frac{\EE\left[(\Dout_v(k))^p\right]}{k^{c_2p}}<\infty.
\end{align*}
\end{lemma}
The proof of Lemma~\ref{lemma:moment} is given in Appendix~\ref{subsec:pf2.2}.

\begin{Remark}\label{rmk:moment}
In the proof of Lemma~\ref{lemma:moment}, if we revise 
the inequality $|V(S_v+k)|\ge |V(k)|$ by $|V(S_v+k)|\ge |V(v+k-1)|$, then we have for $p\ge 1$,
\begin{align}\label{eq:Deg_bound}
\sup_{n\ge 1}
\sup_{v\ge 1}\EE\left(\left(\frac{\Din_v(n)}{(n/v)^{c_1}}\right)^p\right)<\infty,
&\qquad
\sup_{n\ge 1}
\sup_{v\ge 1}\EE\left(\left(\frac{\Dout_v(n)}{(n/v)^{c_2}}\right)^p\right)<\infty.
\end{align}
These results are useful when we study the asymptotic behavior of the proportion 
of reciprocal edges in Section~\ref{subsec:Rn}.
\end{Remark}

The following lemma summarizes the asymptotic properties of $(\Din_v(n), \Dout_v(n))$, from which 
we can derive the limiting behavior of the number of reciprocal edges between two fixed nodes in Section~\ref{subsec:Lij}.
\begin{lemma}\label{lem:prelim}
For $v\ge 1$, there exist random variables $\xi^\text{in}_v$, $\xi^\text{out}_v$ satisfying
$\PP(\xi^\text{in}_v\in (0,\infty))=1=\PP(\xi^\text{out}_v\in (0,\infty))$,
 such that
\begin{align*}
\frac{\Din_v(n)}{n^{c_1}}\convas \xi^\text{in}_v,
\qquad
\frac{\Dout_v(n)}{n^{c_2}}\convas \xi^\text{out}_v.
\end{align*}
\end{lemma}
The proof of Lemma~\ref{lem:prelim} is given in Appendix~\ref{subsec:pf2.3}.

\section{Measuring Reciprocity in a PA Network}\label{sec:recip}

To assess goodness of fit of the directed PA model to a particular dataset, it is useful to evaluate statistics to see if
the empirical values match those from the fitted model. The statistic we focus on here is {\it reciprocity\/}.
If there is a huge discrepancy in the reciprocity measure between the fitted PA model and the given network, then we conclude that variants of the classical PA model should be considered.
In this section, 
we focus on the asymptotic behavior of the number of reciprocal edges between two fixed nodes $i$ and $j$ as well as that of the proportion of reciprocal edges in the entire graph. 

 \begin{figure}[h]
\centering
\includegraphics[scale=.4]{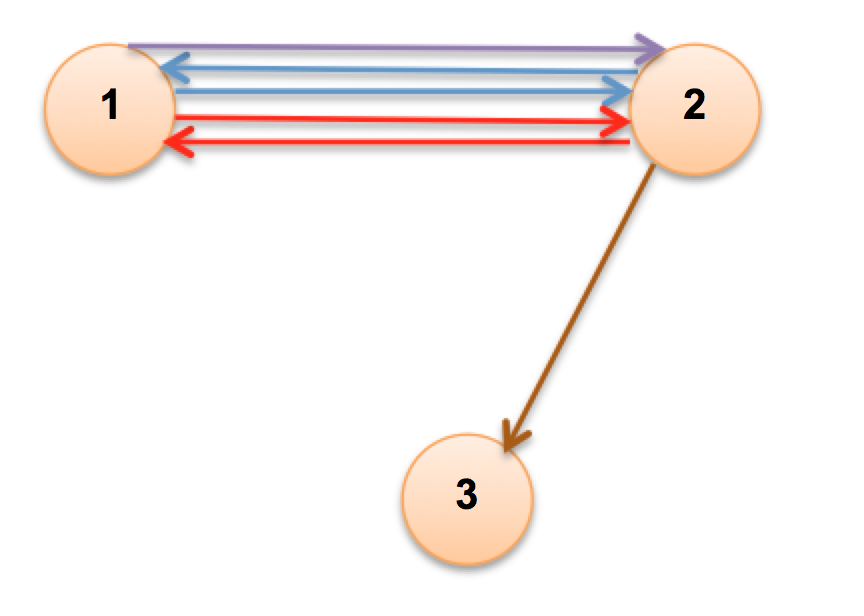}
\caption{A graphical illustration of reciprocal edges with edge set
$E=\{(1,2),(1,2),(1,2),(2,1),(2,1), (2,3)\}$ and $V=\{1,2,3\}$. 
If a pair of reciprocated edges are observed, we label them with the same color. 
Here
we have $R_6 = 4/6 = 0.667$.}\label{fig:g}
\end{figure}

Given an arbitrary graph $G$ with edge set $E$ and node set $V$, we
let $L_{(i,j)}= L_{(i,j)}(G)$ be the number of directed edges $(i,j)$ in graph $G$ for $i,j\in V$. Then
for $|E|=k$, define the \emph{reciprocity coefficient}, $R_k$, as:
\[
R_k = \frac{2}{|E|}\sum_{i,j\in V} \min\left\{L_{(i,j)}, L_{(j,i)}\right\} = \frac{2}{k}\sum_{i,j\in V} \min\left\{L_{(i,j)}, L_{(j,i)}\right\}.
\]
Note that a node pair can be counted more than once.
For example, 
consider the graph given in Figure~\ref{fig:g}, where 
there are $|E|=6$ edges and node set $V=\{1,2,3\}$. 
We distinguish multiple edges between two nodes by different colors,
and if a pair of reciprocated edges are observed, we label the pair with the same color. 
The graph in Figure~\ref{fig:g} contains a pair of blue edges and a pair of red edges, thus giving
$R_6 = 4/6 = 0.667$.
In \verb6R6, we can easily compute the reciprocity coefficient by applying the \verb6dyad_census()6 function in the \verb6igraph6 package to the graph object, and its
 \verb6mut6 value outputs the total number of unordered node pairs $\{i,j\}$ with reciprocal connections $(i,j)$ and $(j,i)$, allowing multiplicity.

Suppose we have a graph $G(n) = (V(n), E(n))$ constructed following the PA rule with parameters $(\alpha,\beta,\gamma,\deltain,\deltaout)$ 
as outlined in Section~\ref{sec:model}. For two nodes $i,j\in V(n)$, write
$L_{(i,j)}(n) = L_{(i,j)}(G(n))$. 
Note that under the PA setup, where $|E(n)|=n+1$, we denote the reciprocity coefficient 
for the PA network $G(n)$ as
\begin{equation}\label{eq:Rn}
R_n^\text{pa} = \frac{2}{n+1}\sum_{i,j\in V(n): i<j} \min\left\{L_{(i,j)}(n), L_{(j,i)}(n)\right\}.
\end{equation}
The definition in \eqref{eq:Rn} excludes self-loops from reciprocal edges.
In Sections~\ref{subsec:Lij} and \ref{subsec:Rn}, we study the asymptotic behavior of $L_{(i,j)}(n)$ for fixed $i,j$, and
$R_n^\text{pa}$ in a PA network $G(n)$, respectively.

\subsection{Reciprocal Edges between Two Fixed Nodes.}\label{subsec:Lij}
In a directed PA network, the total number of reciprocal edges between two fixed nodes, $i,j$, is equal to
\[
L_{i\leftrightarrow j}(n):=2\min\left\{L_{(i,j)}(n), L_{(j,i)}(n)\right\}.
\]
In this section, we will first study the limiting behavior of the number of edges
between two fixed nodes $i,j$, $L_{(i,j)}(n)$, when $n$ is large.
The asymptotics of $L_{i\leftrightarrow j}(n)$ then follow from a continuous mapping argument. With asymptotic results of $L_{(i,j)}(n)$ available, we also give
the behavior of the first time when a reciprocal pair is formed between $i$ and $j$.

\subsubsection{Convergence of $L_{(i,j)}(n)$.}
Main asymptotic results are given in Theorem~\ref{thm:Lij}, and we start 
by presenting a lemma on $A_{ij}(n)$, which is 
 useful for the proof of Theorem~\ref{thm:Lij}.

\begin{lemma}\label{lem:Aij_conv}
Recall the definition of $A_{ij}(n)$ in \eqref{eq:def_Aij}.
For fixed $1\le i<j$, 
\begin{equation}\label{eq:Aij_conv}
\frac{A_{ij}(n)}{n^{c_1+c_2-2}}\longrightarrow \frac{1}{(1+\deltain(1-\beta))(1+\deltaout(1-\beta))}\xi_j^\text{in} \xi_i^\text{out},
\qquad\text{a.s. and in $L_1$}.
\end{equation}
This further gives:
\begin{enumerate}
\item When $c_1+c_2>1$, there exist constants $C_{ij}>0$ such that
\begin{align}
\frac{c_1+c_2-1}{n^{c_1+c_2-1}}\sum_{k=S_j}^{S_j+n} A_{ij}(k) &\longrightarrow C_{ij}\xi_j^\text{in} \xi_i^\text{out}, \qquad\text{a.s. and in $L_1$}.
\label{eq:Aij1a}
\end{align}
\item When $c_1+c_2=1$, there exist constants $C'_{ij}>0$ such that
\begin{align}
\frac{1}{\log n}\sum_{k=S_j}^{S_j+n} A_{ij}(k) &\longrightarrow C'_{ij} \xi_j^\text{in} \xi_i^\text{out},
\qquad\text{a.s. and in $L_1$}. \label{eq:Aij2a}
\end{align}
\item When $c_1+c_2<1$, there exist constants $C''_{ij}>0$ such that
\begin{align}
\frac{1}{n^{c_1+c_2-1}}\sum_{k=n}^{\infty} A_{ij}(k) &\longrightarrow C''_{ij} \xi_j^\text{in} \xi_i^\text{out}, \qquad\text{a.s. and in $L_1$},
\label{eq:Aij3a}
\end{align}
which further implies $\sum_{k\ge 1} \EE(A_{ij}(k)) <\infty$, and $\sum_{k\ge 1} A_{ij}(k) <\infty$ a.s..
\end{enumerate}
In addition, we have similar convergence results for $A_{ji}(n)$ by replacing
$A_{ij}(n)$, $\xi^\text{in}_j$, and $\xi^\text{out}_i$, with 
$A_{ji}(n)$, $\xi^\text{in}_i$, and $\xi^\text{out}_j$, respectively.
\end{lemma}

\begin{proof}
By Lemma~\ref{lem:prelim}, we have
\begin{align*}
\frac{A_{ij}(n)}{n^{c_1+c_2-2}} 
&= \frac{(\Dout_i(n)+\deltaout)(\Din_j(n)+\deltain)}{n^{c_1+c_2}}
\frac{n^{2}}{(n+1+\deltain |V(n)|)(n+1+\deltaout |V(n)|)}\\
&\convas \frac{1}{(1+\deltain(1-\beta))(1+\deltaout(1-\beta))}\xi_j^\text{in} \xi_i^\text{out}.
\end{align*}
Note that once we show
\begin{align}\label{eq:ExpeAij}
\sup_{n\ge 1}\EE\left[\left(\frac{A_{ij}(n)}{n^{c_1+c_2-2}} \right)^2\right]<\infty,
\end{align}
then by \cite[Theorem~4.6.2]{durrett:2019}, $\{A_{ij}(n)/n^{c_1+c_2-2}: n\ge 1\}$ is uniformly integrable, which gives the $L_1$-convergence in \eqref{eq:Aij_conv}.

To prove \eqref{eq:ExpeAij}, we now use the Cauchy-Schwartz inequality to obtain
\begin{align*}
\EE\left[\left(\frac{A_{ij}(n)}{n^{c_1+c_2-2}}\right)^2\right]
&\le \EE\left[\frac{(\Dout_i(n)+\deltaout)^2(\Din_j(n)+\deltain)^2}{n^{2(c_1+c_2)}}\right]\\
&\le \frac{1}{n^{2(c_1+c_2)}}\left(\EE\left[(\Dout_i(n)+\deltaout)^4\right]\EE\left[(\Din_i(n)+\deltain)^4\right]\right)^{1/2}.
\end{align*}
Then by
Lemma~\ref{lemma:moment}, we have
\begin{align*}
&\sup_{n\ge 1}\EE\left[\left(\frac{A_{ij}(n)}{n^{c_1+c_2-2}} \right)^2\right]\\
&\quad\le \sup_{n\ge 1}\frac{\sqrt{\EE\left[(\Dout_i(n)+\deltaout)^4\right]}}{n^{2c_2}}
\times \sup_{n\ge 1}\frac{\sqrt{\EE\left[(\Din_j(n)+\deltain)^4\right]}}{n^{2c_1}}<\infty.
\end{align*}
With \eqref{eq:Aij_conv} established, results in \eqref{eq:Aij1a}--\eqref{eq:Aij3a}
follow directly from 
the Karamata's theorem (cf. \cite[Theorem~2.1]{resnickbook:2007}).
The results for $A_{ji}(n)$ follow from a similar reasoning.
\end{proof}

We now give the asymptotic behavior of $L_{(i,j)}(n)$ in a directed PA model.

\begin{theorem}\label{thm:Lij}
Consider two fixed nodes $1\le i< j$:\\
(i) If $c_1+c_2> 1$, then there exists some random variable $\xi_{ij}$
satisfying $\PP(\xi_{ij}=0) = 0 $, such that as $n\to\infty$,
\begin{equation}\label{eq:L1}
\frac{L_{(i,j)}(n)}{n^{c_1+c_2-1}}\convas \xi_{ij}.
\end{equation}
So for large $n$, the number of $(i,j)$ edges is of order $n^{c_1+c_2-1}$.

\noindent(ii) If $c_1+c_2= 1$, then there exists some random variable $\zeta_{ij}$, 
satisfying $\PP(\zeta_{ij}=0) = 0 $, such that
\begin{equation}\label{eq:L2}
\frac{L_{(i,j)}(n)}{\log n}\convas \zeta_{ij}.
\end{equation}
So for large $n$, the number of $(i,j)$ edges is of order $\log n$.

\noindent(iii)
If $c_1+c_2<1$, then for any $i<j$, there is a last time for an $(i,j)$ edge to form.
Further, as $n\to\infty$,
\begin{equation}\label{eq:L3}
L_{(i,j)}(n)\uparrow L_{(i,j)}(\infty)<\infty,
\qquad\text{a.s.},
\end{equation}
and $L_{(i,j)}(n)-L_{(i,j)}(\infty) = 0$ a.s. for $n$ large.

In addition, we have similar convergence results for $L_{(j,i)}(n)$
by replacing $L_{(i,j)}(n)$, $L_{(i,j)}(\infty)$, $\xi_{ij}$, and $\zeta_{ij}$ with $L_{(j,i)}(n)$,
$L_{(j,i)}(\infty)$, $\xi_{ji}$, and $\zeta_{ji}$, respectively.
\end{theorem}



\begin{proof}
Set 
\[
\Delta_k(i,j) = \ind_{\bigl\{E(k)= E(k-1)\cup \{(i,j)\}\bigr\}},
\]
i.e. $\Delta_k(i,j) = 1$ if a directed edge $(i,j)$ is created from $G(k-1)$ to $G(k)$.
For $1\le i<j$, 
notice that
\begin{equation}\label{eq:Lij}
L_{(i,j)}(S_j+n) = \sum_{k=1}^{S_j+n} \Delta_k(i,j)
=\sum_{k=0}^{n} \Delta_{S_j+k}(i,j).
\end{equation}
For $n\ge 0$,
\begin{align}\label{eq:Lij_beta}
\EE^{\mathcal{F}_{S_j+n}}&\left(\Delta_{S_j+n+1}(i,j)\right)
= \beta A_{ji}(S_j+n),
\end{align}
and 
\begin{align}\label{eq:Lij_gamma}
\EE^{\mathcal{F}_{S_j-1}}&\left(\Delta_{S_j}(i,j)\right)
= \gamma\frac{\Dout_i(S_j-1)+\deltaout}{S_j+\deltaout (j-1)}.
\end{align}

When $c_1+c_2\ge 1$, \eqref{eq:Aij1a} and \eqref{eq:Aij2a} suggest that
 $\sum_{k=0}^{\infty} A_{ji}(S_j+k) = \infty$ a.s.. Then
we apply \cite[Theorem~4.5.5]{durrett:2019} to get:
\begin{equation}\label{eq:Lij_mg}
\frac{L_{(i,j)}(S_j+n) }{\gamma\frac{\Dout_i(S_j-1)+\deltaout}{S_j+\deltaout (j-1)}+\sum_{k=0}^{n-1} \beta A_{ji}(S_j+k) }\convas 1.
\end{equation}
Also, by a similar argument as in \eqref{eq:Aij1a}, we have that when $c_1+c_2>1$,
there exists some constant $\widetilde{C}_{ij}>0$ such that
\begin{align}\label{eq:Lij_order}
&\frac{1}{n^{c_1+c_2-1}}
\sum_{k=0}^{n-1} \beta A_{ji}(S_j+k)
\convas \beta \widetilde{C}_{ij} \xi^\text{out}_i\xi^\text{in}_j .
\end{align}
Then combining \eqref{eq:Lij_mg} with \eqref{eq:Lij_order} gives
\begin{equation}\label{eq:Lij_conv}
\frac{L_{(i,j)}(S_j+n) }{n^{c_1+c_2-1}}\convas 
\beta \widetilde{C}_j\xi^\text{out}_i\xi^\text{in}_j.
\end{equation}
An analogous reasoning is also applicable to the $c_1+c_2=1$ case, 
where the scaling function $n^{c_1+c_2-1}$ is replaced with $\log n$, according to 
\eqref{eq:Aij2a}.

Next, consider the case $c_1+c_2<1$. 
By the corollary in \cite[Chapter IV.6, p. 151]{neveu:1965}, we have a.s.
\begin{align*}
\left\{\sum_{k\ge S_j}\Delta_{k+1}(i,j)<\infty\right\}
&= \left\{\sum_{k\ge S_j}\EE^{\mathcal{F}_k}(\Delta_{k+1}(i,j))<\infty\right\}\\
&= \left\{\sum_{k\ge S_j}\beta A_{ji}(k)<\infty\right\}.
\end{align*}
Using a similar argument as in Lemma~\ref{lem:Aij_conv}(3), we have $\sum_{k\ge S_j} A_{ji}(k)<\infty$ a.s., thus giving
\[
\sum_{k\ge S_j}\Delta_{k+1}(i,j)<\infty\qquad \text{a.s.}.
\]
Hence, 
with probability 1, there is a finite number of $(i,j)$ edges can be formed, and
there exists a last time for an $(i,j)$ edge to form.

\end{proof}

Note that by the definition of $L_{i\leftrightarrow j}(n)$, applying the continuous mapping theorem gives the asymptotic results of 
$L_{i\leftrightarrow j}(n)$, which also depends on the value of $c_1+c_2$:
\begin{enumerate}
\item[(1)] If $c_1+c_2> 1$, then there exists some random variable $\overline{\xi}_{ij}$
satisfying $\PP(\overline{\xi}_{ij}=0) = 0 $, such that as $n\to\infty$,
\[
\frac{L_{i\leftrightarrow j}(n)}{n^{c_1+c_2-1}}\convas \overline{\xi}_{ij}.
\]
So for large $n$, the number of reciprocated edges between $i$ and $j$ is of order $n^{c_1+c_2-1}$.

\item[(2)] If $c_1+c_2= 1$, then there exists some random variable $\overline{\zeta}_{ij}$, 
satisfying $\PP(\overline{\zeta}_{ij}=0) = 0 $, such that
\[
\frac{L_{i\leftrightarrow j}(n)}{\log n}\convas \overline{\zeta}_{ij}.
\]
So for large $n$, the number of reciprocated edges between $i$ and $j$ is of order $\log n$.

\item[(3)]
If $c_1+c_2<1$, then almost surely 
\[
L_{i\leftrightarrow j}(n)\uparrow L_{i\leftrightarrow j}(\infty)<\infty.
\]
\end{enumerate}

Now consider a special case with $\gamma=0$ and $\deltain = \deltaout = \delta>0$, then
\[
c_1+c_2<1 \qquad\Leftrightarrow\qquad \alpha> \frac{1}{1+\delta}.
\]
From Theorem~\ref{thm:Lij}, we see that when the probability of generating a new node in a PA network at each step is too high, 
the node set grows strongly and it is difficult to form reciprocal edges.
Then the number of reciprocal edges between two nodes is finite a.s..
If $\beta = 0$, then for $\deltain, \deltaout>0$, we always have $c_1+c_2<1$, indicating that the number of edges between two fixed nodes is finite a.s. when no edge is added between two existing nodes.

\medskip

\subsubsection{The first time when a reciprocal pair forms.}
Based on Theorem~\ref{thm:Lij}, we present some immediate results with regard to 
the first time when a reciprocal pair of formed between nodes $i$ and $j$.
The $c_1+c_2\ge 1$ scenario is discussed in Corollary~\ref{thm:N0a}, while the
$c_1+c_2<1$ case is analyzed in Proposition~\ref{thm:N0}.

\begin{corollary}\label{thm:N0a}
Let $N_0^{i\leftrightarrow j}$ be the first time when a reciprocal pair of edges $i\leftrightarrow j$ is formed between nodes $i,j$, i.e.
\begin{equation}\label{eq:def_N0}
N_0^{i\leftrightarrow j}:= \inf\left\{n\ge 0: (i,j)\in E(n),\text{and }(j,i)\in E(n)\right\},
\end{equation}
with the convention that $\inf\emptyset = \infty$.
Suppose $1\le i<j$ are fixed, 
and $c_1,c_2$ are as given in \eqref{eq:c1c2}, then
$N_0^{i\leftrightarrow j}<\infty$ a.s. if $c_1+c_2\ge 1$.
 \end{corollary}
 \begin{proof}
We will show that for $c_1+c_2\ge 1$, $\lim_{n\to\infty}\PP(N_0^{i\leftrightarrow j}> S_j+n)=\PP(N_0^{i\leftrightarrow j}=\infty)=0$.
Note that when $c_1+c_2\ge 1$, \eqref{eq:Lij_conv} indicates that $L_{(i,j)}(S_j+n)\convas \infty$, and similarly, $L_{(j,i)}(S_j+n)\convas \infty$.
Therefore, 
\begin{align*}
\PP(N_0^{i\leftrightarrow j}> S_j+n) &\le \PP((i,j)\notin E(S_j+n))
+\PP((j,i)\notin E(S_j+n))\\
&= \PP(L_{(i,j)}(S_j+n)=0)+\PP(L_{(j,i)}(S_j+n)=0)
\to 0,
\end{align*}
as $n\to\infty$.
 \end{proof}


\medskip

When $c_1+c_2<1$, if we have $\EE(L_{(i,j)}(\infty))<1$,
then 
\begin{align*}
\PP(L_{(i,j)}(\infty) <1)>0,
\end{align*}
which implies $\PP(L_{(i,j)}(\infty) =0)>0$, and
$\PP(N_0^{i\leftrightarrow j} =\infty)>0$.
Note that
\[
\EE(L_{(i,j)}(\infty)) = \gamma\EE\left(\frac{\Dout_i(S_j-1)+\deltaout}{S_j+\deltaout (j-1)}\right)+
\beta \sum_{k=0}^\infty \EE\left(A_{ji}(S_j+k)\right),
\]
and by \eqref{eq:Deg_bound}, there exists some constants ${K}_1, {K}_2>0$,  such that
for $n\ge i\ge 1$,
\begin{align*}
\EE\bigl((\Din_i(n))^2\bigr)\le K_1(n/i)^{2c_1},\qquad
\EE\bigl((\Dout_i(n))^2\bigr)\le K_2(n/i)^{2c_2}.
\end{align*}
Then applying the Cauchy-Schwartz inequality gives that for $n\ge j>i\ge 1$,
\begin{align*}
\EE\bigl(A_{ji}(n)\bigr)&\le \left[\EE\left(\left(\frac{\Din_i(n)+\deltain}{n+1+\deltain |V(n)|}\right)^2\right)\right]^{1/2}
\left[\EE\left(\left(\frac{\Dout_j(n)+\deltaout}{n+1+\deltaout |V(n)|}\right)^2\right)\right]^{1/2}\\
&\le \frac{1}{n^2}\left[\EE\bigl((\Din_i(n)+\deltain)^2\bigr)\EE\bigl((\Dout_j(n)+\deltaout)^2\bigr)\right]^{1/2}\\
&\le (K_1K_2)^{1/2}\, \frac{n^{c_1+c_2-2}}{i^{c_1}j^{c_2}} =: K'\frac{n^{c_1+c_2-2}}{i^{c_1}j^{c_2}}.
\end{align*}
Therefore, we have
\begin{align*}
\sum_{k=0}^\infty \EE\left(A_{ji}(S_j+k)\right)
&\le {K'}
\sum_{k=j}^\infty \frac{k^{c_1+c_2-2}}{i^{c_1}j^{c_2}}
\le \frac{K'}{1-c_1-c_2}
i^{-c_1}j^{c_1-1},
\end{align*}
which gives
\begin{align}\label{eq:Lij_inf}
\EE(L_{(i,j)}(\infty)) \le \gamma + \frac{\beta K'}{1-c_1-c_2}
i^{-c_1}j^{c_1-1}.
\end{align}
Since $c_1-1<0$, then for $j$ sufficiently large, we have $\EE(L_{(i,j)}(\infty))<1$, thus giving 
$\PP(N_0^{i\leftrightarrow j} =\infty)>0$. In other words, when $c_1+c_2<1$,
it is possible to have zero pair of reciprocal edges between two fixed nodes $i,j$, if $j$ 
is created at a late stage of network evolution. 

Equation \eqref{eq:Lij_inf} also suggests that for arbitrarily chosen $i,j$, $\EE(L_{(i,j)}(\infty))<1$ if $\beta$ is small enough. Hence, if few edges are created between existing nodes, 
it is possible for two fixed nodes to never form a reciprocal pair of edges.
In particular, when $\beta = 0$, there is no reciprocal pair of edges, i.e.
$\PP(N_0^{i\leftrightarrow j} =\infty)=1$, for all fixed $i,j$. 

In the following proposition, we give the asymptotic behavior of 
$\sup_{j\ge n\epsilon} \PP(N_0^{i\leftrightarrow j}\le n)$ in the $c_1+c_2<1$ case.
\begin{prop}\label{thm:N0}
If $c_1+c_2<1$, then for $i\ge 1$, $\epsilon>0$, 
$\sup_{j\ge n\epsilon} \PP(N_0^{i\leftrightarrow j}\le n) \to 0$, as $n\to\infty$.
\end{prop}
\begin{proof}
Applying the union bound to $\PP(N_0^{i\leftrightarrow j}\le S_{j}+n)$ gives:
\begin{align*}
\PP(N_0^{i\leftrightarrow j}\le S_{j}+ n)
&\le \sum_{k=1}^n \PP\bigl(E(S_j+k)= E(S_j+k-1)\cup \{(i,j)\}\bigr)\\
&\qquad + \sum_{k=1}^n\PP\bigl(E(S_j+k)= E(S_j+k-1)\cup \{(j,i)\}\bigr)\\
&= \EE\left(\sum_{k=0}^{n-1} \beta A_{ji}(S_{j}+k)\right)
 + \EE\left(\sum_{k=0}^{n-1} \beta A_{ij}(S_{j}+k)\right)\\
 &\le \EE\left(\sum_{k=j}^{\infty} \beta A_{ji}(k)\right)
 + \EE\left(\sum_{k=j}^{\infty} \beta A_{ij}(k)\right).
\end{align*} 
By \eqref{eq:Aij3a}, we see that for $\epsilon>0$,
\[
\sup_{j\ge n\epsilon}\EE\left(\sum_{k=j}^{\infty} A_{ji}(k)\right) \to 0,
\qquad \sup_{j\ge n\epsilon}\EE\left(\sum_{k=j}^{\infty} A_{ij}(k)\right) \to 0,
\]
as $n\to\infty$, which gives
$$\sup_{j\ge n\epsilon}\PP(N_0^{i\leftrightarrow j}\le n) \le \sup_{j\ge n\epsilon}\PP(N_0^{i\leftrightarrow j}\le S_j+n)  \to 0.
$$
\end{proof}

\subsection{Reciprocity in the Entire Graph.}\label{subsec:Rn}

Then we consider the proportion of reciprocal edges in the entire PA network, $R_n^\text{pa} $, and 
the next theorem specifies the asymptotic behavior of $R_n$ for $0< c_1+c_2< 5/3$.
\begin{theorem}\label{thm:Mn}
Suppose $R_n^\text{pa}$ is as defined in \eqref{eq:Rn}. Then
for $0< c_1+c_2< 5/3$, we have
\[
R_n^\text{pa}  \stackrel{p}{\longrightarrow} 0,\qquad n\to\infty.
\]
\end{theorem}

Results in Theorem~\ref{thm:Mn} show that for certain combination of model parameters
such that $c_1+c_2<5/3$, it is likely to have a small reciprocity coefficient, $R_n^\text{pa} $, provided that the number of edges in the PA network is large.
In particular, when $c_1+c_2\le 1$, i.e. the second and third cases in Theorem~\ref{thm:Lij}, we have $R_n^\text{pa}\stackrel{p}{\longrightarrow} 0$.

\begin{proof}
Note that it suffices to show $\EE(R_n^\text{pa} )\to 0$ for $0< c_1+c_2< 5/3$, as $n\to \infty$.
Recall that $$\Delta_k(i,j) = \ind_{\bigl\{E(k)= E(k-1)\cup \{(i,j)\}\bigr\}},$$
then $\Delta_k(i,j)\ind_{\{(j,i)\in E(k-1)\}}$ denotes the event that
from $G(k-1)$ to $G(k)$, an edge $(i,j)$ is created when $(j,i)$ already exists in $G(k-1)$.
By the definition of $R_n^\text{pa} $, we have
\begin{align*}
R_n^\text{pa}  &\le \frac{2}{n+1}\sum_{j=1}^n \sum_{i<j} \sum_{k= S_j}^n 
\Delta_k(i,j)\ind_{\{(j,i)\in E(k-1)\}}\\
&\qquad+\frac{2}{n+1}\sum_{j=1}^n \sum_{i<j} \sum_{k= S_j}^n 
\Delta_k(i,j)\ind_{\{(i,j)\in E(k-1)\}}\\
&= \frac{2}{n+1}\sum_{j=1}^n \sum_{i<j} \sum_{k= S_j+1}^n 
\Delta_k(i,j)\ind_{\{(j,i)\in E(k-1)\}}\\
&\qquad+\frac{2}{n+1}\sum_{j=1}^n \sum_{i<j} \sum_{k= S_j+1}^n 
\Delta_k(i,j)\ind_{\{(i,j)\in E(k-1)\}}\\
&=: Q_1(n) + Q_2(n).
\end{align*}
For $Q_1(n)$, we have
\begin{align}
\EE(Q_1(n)) &=
\frac{2}{n+1}\sum_{j=1}^n \sum_{i<j} 
\EE\left(\sum_{k= S_j+1}^n \Delta_k(i,j)\ind_{\{(j,i)\in E(k-1)\}}\right)\nonumber\\
&= \frac{2}{n+1}\sum_{j=1}^n \sum_{i<j} 
\EE\left(\sum_{k= S_j+1}^n \EE^{\mathcal{F}_{k-1}}\left(\Delta_k(i,j)\right)\ind_{\{(j,i)\in E(k-1)\}}\right)\nonumber\\
&= \frac{2}{n+1}\sum_{j=1}^n \sum_{i<j} 
\EE\left(\sum_{k= S_j+1}^n \beta A_{ji}(k-1)\ind_{\{(j,i)\in E(k-1)\}}\right).\label{eq:Q1}
\end{align}
Since $S_j\ge j-1$ for $j\ge 2$, then \eqref{eq:Q1} implies
\begin{align}
\EE(Q_1(n)) &\le  \frac{2}{n+1}\sum_{j=2}^n \sum_{i=1}^{j-1} 
\sum_{k= j}^n \EE\left(A_{ji}(k-1)\ind_{\{(j,i)\in E(k-1), k\ge S_j+1\}}\right).
\label{eq:Q1_bound}
\end{align}
By the Cauchy-Schwartz inequality, we have
\begin{align}
\EE&\left(A_{ji}(k-1)\ind_{\{(j,i)\in E(k-1), k\ge S_j+1\}}\right)\nonumber\\
&\le \left[\EE\left(A^2_{ji}(k-1)\right)\right]^{1/2} \left[\PP((j,i)\in E(k-1), k\ge S_j+1)\right]^{1/2}\nonumber\\
&\le \left[\EE\left(A^2_{ji}(k-1)\right)\right]^{1/2}\left[\sum_{l=j}^{k-1}\PP\left(E(l)= E(l-1)\cup \{(i,j)\}, l\ge S_j+1\right)\right]^{1/2}\nonumber\\
&\le \left[\EE\left(A^2_{ji}(k-1)\right)\right]^{1/2}\left[\sum_{l=j}^{k-1}\EE\left(A_{ij}(l)\right)+\alpha
\EE\left(\frac{\Din_i(S_j-1)+\deltain}{S_j+\deltain (j-1)}\right)\right]^{1/2}.
\label{eq:Q1_CSbound}
\end{align}

When $c_1+c_2>1$, Lemma~\ref{lemma:moment} and \eqref{eq:Deg_bound} together imply that there exist constants $M_1$, $M_2$, $M_3>0$ such that for $i<j\le k\le n$,
\[
\EE\left(A^2_{ji}(k-1)\right) \le M_1\frac{k^{2(c_1+c_2-2)}}{i^{2c_2}j^{2c_1}},
\quad \EE\left(\frac{\Din_i(S_j-1)+\deltain}{S_j+\deltain (j-1)}\right)\le M_3\frac{j^{c_1-1}}{i^{c_1}},
\]
and that for $j\le l\le k-1$,
\[
 \EE\left(A_{ij}(l)\right) \le M_2\frac{l^{c_1+c_2-2}}{i^{c_1}j^{c_2}}.
\]
Therefore, when $c_1+c_2>1$, we have
\begin{align*}
\EE&\left(A_{ji}(k-1)\ind_{\{(j,i)\in E(k-1), k\ge S_j+1\}}\right)\\
&\le \sqrt{M_1}\frac{k^{c_1+c_2-2}}{i^{c_2}j^{c_1}} \left(\frac{{M_2}}{c_1+c_2-1}\frac{k^{c_1+c_2-1}}{i^{c_1}j^{c_2}}+\alpha M_3 j^{c_1-1}/i^{c_1}\right)^{1/2},
\end{align*}
and since $k^{c_1+c_2-1}/(i^{c_1}j^{c_2})\ge j^{c_1-1}/i^{c_1}$ for $k\ge j$,
there exists some constant $M>0$, 
\begin{align}\label{eq:Aji_bound}
\EE&\left(A_{ji}(k-1)\ind_{\{(j,i)\in E(k-1), k\ge S_j+1\}}\right)
\le M \frac{k^{\frac{3}{2}(c_1+c_2)-\frac{5}{2}}}{i^{c_1/2+c_2}j^{c_1+c_2/2}}.
\end{align}
Then \eqref{eq:Q1_bound} leads to
\begin{align*}
\EE(Q_1(n)) &\le \frac{2M}{n+1}\sum_{j=2}^n \sum_{i=1}^{j-1} 
\sum_{k= j}^n \frac{k^{\frac{3}{2}(c_1+c_2)-\frac{5}{2}}}{i^{c_1/2+c_2}j^{c_1+c_2/2}}\\
&\le \frac{2M}{3/2(c_1+c_2-1)}n^{\frac{3}{2}(c_1+c_2)-\frac{5}{2}}
\sum_{j=2}^n \sum_{i=1}^{j-1} i^{-(c_1/2+c_2)} j^{-(c_1+c_2/2)}.
\end{align*}
If $c_1/2+c_2>1$, $c_1+c_2/2>1$, and $c_1+c_2<5/3$, then 
\begin{align*}
\EE(Q_1(n)) &\le \frac{2M}{3/2(c_1+c_2-1)}n^{\frac{3}{2}(c_1+c_2)-\frac{5}{2}}
\sum_{i=1}^{\infty} i^{-(c_1/2+c_2)} \sum_{j=2}^\infty  j^{-(c_1+c_2/2)} \to 0.
\end{align*}
If $c_1/2+c_2< 1$, $c_1+c_2/2< 1$, then
\begin{align*}
\EE(Q_1(n)) &\le \frac{2M}{3/2(c_1+c_2-1)} n^{\frac{3}{2}(c_1+c_2)-\frac{5}{2}}\frac{n^{2-3/2(c_1+c_2)}}{(1-c_1/2-c_2)(1-c_1-c_2/2)}\\
&=\frac{2M n^{-1/2}}{3/2(c_1+c_2-1)(1-c_1/2-c_2)(1-c_1-c_2/2)} \to 0.
\end{align*}
If $c_1/2+c_2<1$, $c_1+c_2/2>1$, and $c_1+c_2<5/3$, then 
\begin{align*}
\EE(Q_1(n)) &\le \frac{2M}{3/2(c_1+c_2-1)} n^{\frac{3}{2}(c_1+c_2)-\frac{5}{2}}
\frac{1}{1-(c_1/2+c_2)}\sum_{j=2}^n j^{1-(c_1/2+c_2)}j^{-1}\\
&\le \frac{2M n^{c_1+c_2/2-3/2}}{3/2(c_1+c_2-1)(1-(c_1/2+c_2))^2}\to 0,
\end{align*}
as $c_1+c_2/2<1+1/2=3/2$. Similarly,  $\EE(Q_2(n)) \to 0$, when 
$c_1/2+c_2>1$, $c_1+c_2/2<1$, and $c_1+c_2<5/3$.
The proof machinery also applies to the case where either $c_1/2+c_2=1$
or $c_1+c_2/2=1$, and $c_1+c_2<5/3$, which gives the conclusion that for 
$1<c_1+c_2<5/3$, 
$\EE(Q_1(n)) \to 0$.
Following the same reasoning, we have $\EE(Q_2(n)) \to 0$, for $1<c_1+c_2<5/3$, 
thus implying $R_n^\text{pa} \stackrel{p}{\longrightarrow} 0$ for $1<c_1+c_2<5/3$.

When $c_1+c_2=1$, we revise the bound in \eqref{eq:Aji_bound} to get: for some 
constant $\widetilde{M}>0$,
\begin{align*}
\EE&\left(A_{ji}(k-1)\ind_{\{(j,i)\in E(k-1), \, k\ge S_j+1\}}\right)
\le \widetilde{M} \frac{k^{-1}(\log k)^{1/2}}{i^{c_1/2+c_2}j^{c_1+c_2/2}}.
\end{align*}
Then we have
\begin{align*}
\EE(Q_1(n)) &\le \frac{2\widetilde{M}}{n}(\log n)^{3/2}\sum_{j=2}^n\sum_{i=1}^{j-1} i^{-(c_1/2+c_2)}j^{-(c_1+c_2/2)}\\
&\le \frac{2\widetilde{M}}{(1-c_1/2-c_2)(1-c_1-c_2/2)}\frac{(\log n)^{3/2}}{n^{1/2}}\to 0.
\end{align*}
Meanwhile, for some constant $\widetilde{M}'>0$, we have
\begin{align*}
\EE(Q_2(n)) &\le \frac{2\widetilde{M}'}{n}(\log n)^{3/2}\sum_{j=2}^n\sum_{i=1}^{j-1} i^{-(c_2/2+c_1)}j^{-(c_2+c_1/2)}\\
&\le \frac{2\widetilde{M}'}{(1-c_1/2-c_2)(1-c_1-c_2/2)}\frac{(\log n)^{3/2}}{n^{1/2}}\to 0.
\end{align*}
Hence, we have $R_n^\text{pa} \stackrel{p}{\longrightarrow}0$ when $c_1+c_2=1$.

When $c_1+c_2<1$, the bound in \eqref{eq:Q1_CSbound} implies that there exists
some constant $\bar M>0$ such that
\begin{align}
\EE\left(A_{ji}(k-1)\ind_{\{(j,i)\in E(k-1), k\ge S_j+1\}}\right)
&\le \bar M \frac{k^{c_1+c_2-2}}{i^{c_2}j^{c_1}}\left(\frac{j^{c_1+c_2-1}}{i^{c_1}j^{c_2}}\right)^{1/2}\nonumber\\
&= \bar M \frac{k^{c_1+c_2-2}}{i^{c_1/2+c_2}j^{c_1/2+1/2}},
\label{eq:Aji_case3}
\end{align}
which gives 
\begin{align*}
\EE(Q_1(n))&\le \frac{2\bar M}{(1-c_1-c_2)(1-c_1/2-c_2)}n^{-1/2}\to 0.
\end{align*}
Similar reasoning also gives $\EE(Q_2(n))\to 0$ when $c_1+c_2<1$, thus giving
$\EE(R_n^\text{pa} )\to 0$ and completing the proof of the theorem.

\end{proof}
\begin{Remark}\label{rmk:Rn}
{\rm

\noindent(i) Recall the definition of $c_1$ and $c_2$ in \eqref{eq:c1c2}, and if $\beta\le 2/3$, then $c_1+c_2<1+\beta\le 5/3$. Theorem~\ref{thm:Mn} suggests that if the proportion of edges added between two existing nodes is less than $2/3$ and $n$ is sufficiently large, 
the corresponding PA network will have $R_n^\text{pa} $ close to 0.

\noindent(ii) Note also that 
\[
\ind_{\{N_0^{i\leftrightarrow j}\le n\}}
\le \sum_{k=1}^n \Delta_k(i,j)\ind_{\{(j,i)\in E(k-1)\}}
+\sum_{k=1}^n \Delta_k(j,i)\ind_{\{(i,j)\in E(k-1)\}}.
\]
Hence, when $c_1+2c_2<1$, applying the bound in \eqref{eq:Aji_case3} gives: for fixed $i$,
there exists some constant $\tilde{M}>0$ such that
\[
\lim_{n\to\infty}\EE\left(\sum_{j\in V(n)}\ind_{\{N_0^{i\leftrightarrow j}\le n\}}\right) 
\le \tilde{M}i^{-(c_1/2+c_2)}\sum_{j=1}^\infty j^{c_1/2+c_2-3/2}<\infty.
\]
This indicates when $c_1+2c_2<1$, a fixed node $i$ can form a reciprocal pair of edges only with
finitely many nodes. }
\end{Remark}

\subsection{Simulation for $c_1+c_2\ge 5/3$.}

We are left with the asymptotic behavior of $R_n^\text{pa} $ for $c_1+c_2\in [5/3, 1+\beta)$, provided that
$\beta\in (2/3,1)$.
For comparison, we choose three sets of parameters:
\begin{align*}
\boldsymbol{\theta}_1&= (\alpha,\beta,\gamma,\deltain,\deltaout) = (0.1,0.8,0.1,2,1),\\
\boldsymbol{\theta}_2& = (0.1,0.8,0.1,.4,.4),\qquad\text{and}\qquad
\boldsymbol{\theta}_3 = (0.05,0.9,0.05,1,1),
\end{align*}
such that values of $c_1+c_2$ are 
equal to 1.393, 1.667, and 1.727, respectively.
In other words, the three sets of parameter values correspond to the cases where $c_1+c_2$ are
strictly smaller than, equal to, and strictly greater than $5/3$, respectively. For each $\boldsymbol{\theta}_i$, $i=1,2,3$, we simulate 1000 replications of the directed PA network with $10^5$ edges, and compute the value of $R_n^\text{pa} $ for each replication.

\begin{figure}[h]
\centering
\includegraphics[scale=.6]{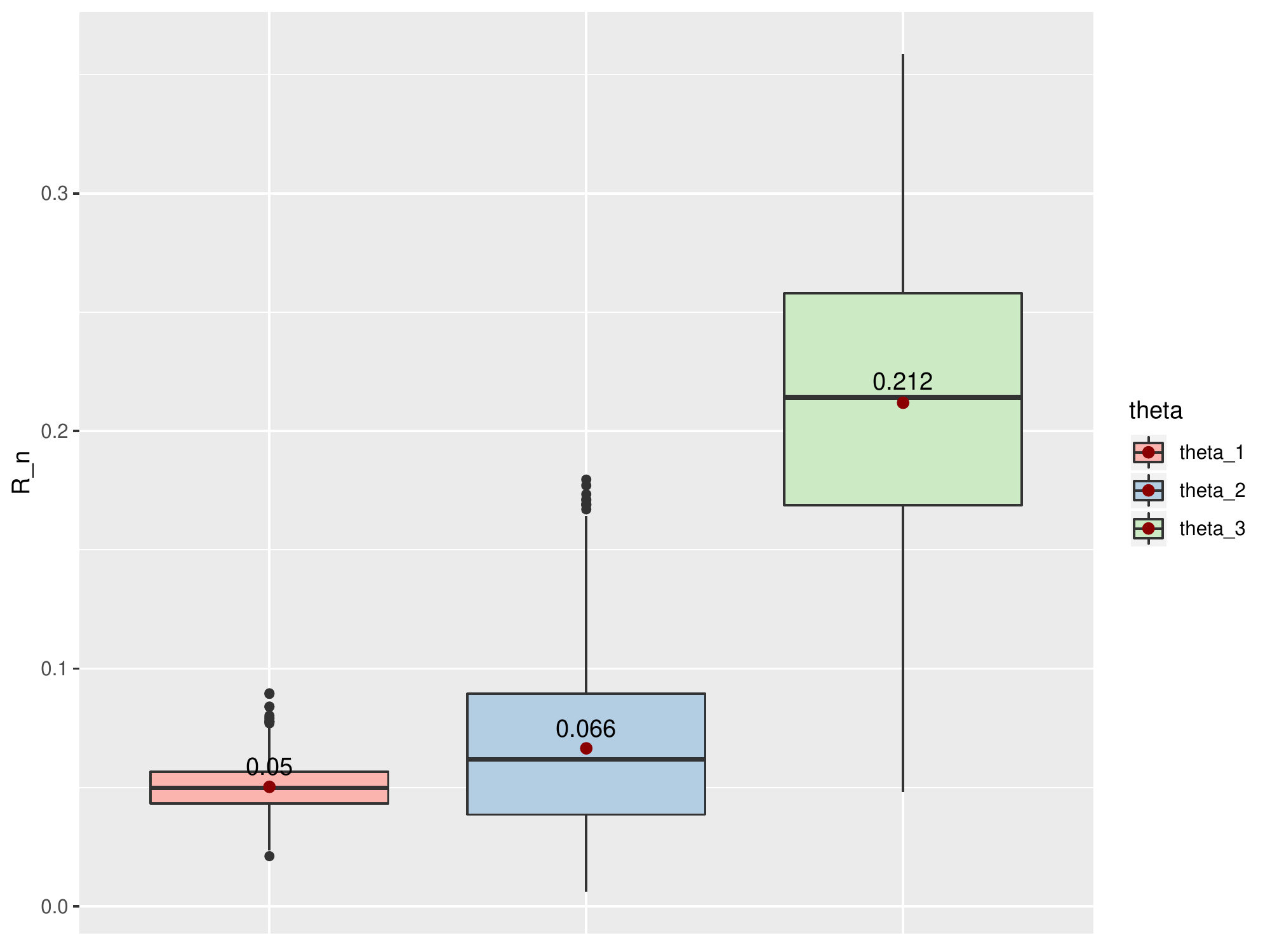}
\caption{Boxplots of $R_n^\text{pa} $ for directed PA models simulated using $\boldsymbol{\theta}_i$, $i=1,2,3$. The red dots represent the averaged empirical $R_n^\text{pa} $ for each $\boldsymbol{\theta}_i$, $i=1,2,3$.}\label{fig:boxplot}
\end{figure}

Numerical results are summarized as boxplots in Figure~\ref{fig:boxplot}. For each boxplot, we use the dark red dots to mark the corresponding averaged empirical $R_n^\text{pa} $.
Under $\boldsymbol{\theta}_1$, all 1000 empirical $R_n^\text{pa} $ values are close to 0, with a maximum of 0.090 and a minimum of 0.021. The empirical $R_n^\text{pa} $ values under $\boldsymbol{\theta}_2$ and $\boldsymbol{\theta}_3$ are more variable, but both have a higher mean than the $\boldsymbol{\theta}_1$ case. This simulation experiment confirms that the asymptotic behavior of $R_n^\text{pa} $ in a directed PA model depends on the value of $c_1+c_2$. Meanwhile, when $c_1+c_2\ge 5/3$, the value of $R_n^\text{pa} $ may not necessarily concentrate around a specific value, but vary over a certain range.

\section{Discussion}\label{sec:discuss}

Suppose that we are given a scale-free network with a large proportion of reciprocal edges, e.g. Facebook wall posts \cite{facebook:2009}, Twitter \cite{java:etal:2007}, Google$+$ \cite{googleplus:2012}, and Flickr \cite{flickr:2009,social:2007}.
When fitting a directed PA model to such a dataset using inference methods developed in \cite{wan:wang:davis:resnick:2017, wan:wang:davis:resnick:2017b}, there is no guarantee that the 
calibrated model also has a large $R^\text{pa}_n$.
In fact, estimated $\hat{c}_1$ and $\hat{c}_2$ do not necessarily satisfy $\hat{c}_1+\hat{c}_2\ge 5/3$. If we have $\hat{c}_1+\hat{c}_2<5/3$ in the calibrated model, then by Theorem~\ref{thm:Mn}, the corresponding $R_n^\text{pa}$ is close to 0,
which differs from the feature of high reciprocity in the given dataset.
This flags modeling error and suggests considering alternative models or 
variants of the classical PA network. 
For instance, once a directed edge $(i,j)$ is created following the PA rule, we may add
a reciprocal edge $(j,i)$ with probability $\rho\in (0,1)$. 
The study in \cite{chengetal:2011} also provides other features that can be employed to predict reciprocal edges, and we will defer the analysis of these variants of directed PA models as future research.

\appendix
\section{Proofs in Section~\ref{sec:prelim}}\label{sec:pf}
In this section, we give proofs of lemmas in Section~\ref{sec:prelim}.
\subsection{Proof of Lemma~\ref{lem:fil}}\label{subsec:pf2.1}
Note that the right hand sides of \eqref{eq:deg_1} and \eqref{eq:deg_p} are both $\mathcal{F}_{S_v+k}$-measurable.
We show the results for $p\ge 2$, and the case $p=1$ follows by a similar argument.
Let $F\in \mathcal{F}_{S_v+k}$, and
\begin{align}
&\int_F \EE^{\mathcal{F}_{S_v+k}}\left(\left(\Din_v(S_v+k+1)+\deltain\right)^p\right)\dd \PP
\nonumber\\
&= \int_F \left(\Din_v(S_v+k+1)+\deltain\right)^p\dd \PP\nonumber\\
&= \sum_{l\ge k}\int_{F\cap \{S_v+k=l\}} \left(\Din_v(l+1)+\deltain\right)^p\dd \PP,\nonumber\\
\intertext{and since $\Din_v(l+1) = \Din_v(l) + \ind_{\{\text{Node $v$ is chosen at step $l+1$}\}}
=: \Din_v(l) + \Delta_v(l+1)$,}
&= \sum_{l\ge k}\int_{F\cap \{S_v+k=l\}} \left(\Din_v(l)+\deltain+ \Delta_v(l+1)\right)^p\dd \PP\nonumber\\
&= \sum_{l\ge k}\int_{F\cap \{S_v+k=l\}} \left(\left(\Din_v(l)+\deltain\right)^p
+ \sum_{r=1}^p {p \choose r}(\Din_v(l)+\deltain)^{p-r}\Delta_v(l+1)\right)\dd \PP.
\label{eq:Lem2.1}
\end{align}
Since $F\cap \{S_v+k=l\}\in \mathcal{F}_l$, then the quantity in \eqref{eq:Lem2.1} is equal to
\begin{align*}
&\sum_{l\ge k}\int_{F\cap \{S_v+k=l\}} \left(\Din_v(l)+\deltain\right)^p\dd\PP\\
&\qquad+ \sum_{l\ge k}\int_{F\cap \{S_v+k=l\}} p\left(\Din_v(l)+\deltain\right)^{p-1}\EE^{\mathcal{F}_l}\left(\Delta_v(l+1)\right)\dd \PP\\
&\qquad + \sum_{r=2}^p \sum_{l\ge k}\int_{F\cap \{S_v+k=l\}} {p \choose r}(\Din_v(l)+\deltain)^{p-r}\EE^{\mathcal{F}_l}\left(\Delta_v(l+1)\right)\dd \PP.
\end{align*}
Note also that $\EE^{\mathcal{F}_l}\left(\Delta_v(l+1)\right)=(\alpha+\beta)(\Din_v(l)+\deltain)/(l+1+\deltain |V(l)|)$, so we have
\begin{align*}
&\int_F \EE^{\mathcal{F}_{S_v+k}}\left(\left(\Din_v(S_v+k+1)+\deltain\right)^p\right)\dd \PP
\nonumber\\
&= \int_F(\Din_v(S_v+k)+\deltain)^p\left(1+p\frac{\alpha+\beta}{S_v+k+1+\deltain|V(S_v+k)|}\right)\dd\PP\\
&\qquad+\int_F\frac{\alpha+\beta}{S_v+k+1+\deltain|V(S_v+k)|}\sum_{r=2}^p{p \choose r}(\Din_v(S_v+k)+\deltain)^{p-r+1}\dd \PP.
\end{align*}

\subsection{Proof of Lemma~\ref{lemma:moment}}\label{subsec:pf2.2}
For $p=1$, we see from \eqref{eq:deg_1} that
\begin{align}
\EE&\left(\Din_v(S_v+k+1)+\deltain\right) \nonumber\\
&=\EE\left((\Din_v(S_v+k)+\deltain)\left(1+\frac{\alpha+\beta}{S_v+k+1+\deltain |V(S_v+k)|}\right)\right),\nonumber\\
\intertext{and since $S_v\ge 0$ and $|V(S_v+k)|\ge |V(k)|$, then}
&\le \EE\left((\Din_v(S_v+k)+\deltain)\left(1+\frac{\alpha+\beta}{k+1+\deltain |V(k)|}\right)\right),\nonumber\\
&\le \EE\left(\Din_v(S_v+k)+\deltain\right)\left(1+\frac{c_1}{k}\right)\nonumber\\
&\quad+ \EE\left(\left(\Din_v(S_v+k)+\deltain\right)\frac{(\alpha+\beta)\deltain\left||V(k)|-(1-\beta)k\right|}{(k+\deltain |V(k)|)(1+\deltain (1-\beta))k}\right)\nonumber\\
\intertext{and as $\alpha+\beta\le 1$, we have}
&\le \EE\left(\Din_v(S_v+k)+\deltain\right)\left(1+\frac{c_1}{k}\right)\nonumber\\
&\quad+ \EE\left(\left(\Din_v(S_v+k)+\deltain\right)\frac{\deltain\left||V(k)|-(1-\beta)k\right|}{(k+\deltain |V(k)|)(1+\deltain (1-\beta))k}\right)\nonumber\\
&=: H^{(1)}_v(k) + H^{(2)}_v(k).\label{eq:deg1_diff}
\end{align}
Applying the Chernoff bound, we obtain
\begin{equation}\label{eq:chernoff}
\PP\left(\Bigl|\big|V(k)\big| - (1-\beta)k\Bigr|\ge \sqrt{12(1-\beta) k\log k}\right)\le \frac{2}{k^4},
\end{equation}
and rewrite the term $H^{(2)}_v(k)$ in \eqref{eq:deg1_diff} as
\begin{align}
&H^{(2)}_v(k)\nonumber\\
&= \EE\left(\left(\Din_v(S_v+k)+\deltain\right)\frac{\deltain\left||V(k)|-(1-\beta)k\right|}{(k+\deltain |V(k)|)(1+\deltain (1-\beta))k}
\boldsymbol{1}_{\left\{\left||V(k)| - (1-\beta)k\right|\le \sqrt{12(1-\beta) k\log k}\right\}}\right)
\nonumber\\
&+ \EE\left(\left(\Din_v(S_v+k)+\deltain\right)\frac{\deltain\left||V(k)|-(1-\beta)k\right|}{(k+\deltain |V(k)|)(1+\deltain (1-\beta))k}
\boldsymbol{1}_{\left\{\left||V(k)| - (1-\beta)k\right|> \sqrt{12(1-\beta) k\log k}\right\}}\right).\nonumber
\end{align}
Since $\Din_v(S_v+k)\le k+1$ and $\bigl||V(k)|-(1-\beta)k\bigr|\le k+1$, then the forgoing term is bounded by
\begin{align}
& \EE\left(\Din_v(S_v+k)+\deltain\right)\frac{\deltain\sqrt{12(1-\beta) k\log k}}{k^2}
+\frac{\deltain (k+1+\deltain)(k+1)}{(1+\deltain (1-\beta))k^2}\frac{2}{k^4}\nonumber\\
&\le \EE\left(\Din_v(S_v+k)+\deltain\right)\frac{\deltain\sqrt{12k\log k}}{k^2}
+\frac{2\deltain(k+1+\deltain)(k+1)}{k^6}.\label{eq:deg1_bound}
\end{align}
Combining the bound in \eqref{eq:deg1_bound} with \eqref{eq:deg1_diff} gives
\begin{align*}
\EE&\left(\Din_v(S_v+k+1)+\deltain\right) \\
&\le\EE\left(\Din_v(S_v+k)+\deltain\right)\left(1+\frac{c_1}{k}\right)\\
 &\quad+ \EE\left(\left(\Din_v(S_v+k)+\deltain\right)\frac{\deltain\left||V(k)|-(1-\beta)k\right|}{(k+\deltain |V(k)|)(1+\deltain (1-\beta))k}\right)\\
&\le \EE\left(\Din_v(S_v+k)+\deltain\right)\left(1+\frac{c_1}{k} + \frac{\deltain\sqrt{12 k\log k}}{k^2}\right)\\
&\quad+ \frac{2\deltain(k+1+\deltain)(k+1)}{k^6}.
\end{align*}

Recursively applying the inequality above $k$ times, 
we have:
\begin{align}
\EE&\left(\Din_v(S_v+k+1)+\deltain\right)\nonumber\\
&\le\EE\left(\Din_v(S_v+k)+\deltain\right)\left(1+\frac{c_1}{k} + \frac{\deltain\sqrt{12 k\log k}}{k^2}\right)\nonumber\\
&\qquad+ \frac{2\deltain(k+1+\deltain)(k+1)}{k^6}\nonumber\\
&\le\ldots\le 
\EE(\Din_v(S_v)+\deltain)\prod_{l=1}^k \left(1+\frac{c_1}{l}+\frac{\deltain\sqrt{12 l\log l}}{l^2}\right)\nonumber\\
&+ 2\deltain\sum_{l=1}^k \frac{(l+1+\deltain)(l+1)}{l^6}\prod_{s=l+1}^k \left(1+\frac{c_1}{s}+\frac{\deltain\sqrt{12 s\log s}}{s^2}\right).\label{eq:moment_p1}
\end{align}
Here, $\EE(\Din_v(S_v)+\deltain)=\alpha\deltain +\gamma(1+\deltain)$, depending on whether
the $\alpha$- or $\gamma$-scenario occurs.
Note that there exists a constant $M>0$ such that
\begin{align}\label{eq:bound_D1}
\prod_{l=1}^k \left(1+\frac{c_1}{l}+\frac{\deltain\sqrt{12 l\log l}}{l^2}\right)
\le \exp\left\{\sum_{l=1}^k\left(\frac{c_1}{l}+\frac{\deltain\sqrt{12 l\log l}}{l^2}\right)\right\}
\le M k^{c_1},
\end{align}
and it follows from \eqref{eq:moment_p1} that 
\[
\sup_{k\ge 1}\frac{\EE\left[\Din_v(S_v+k)\right]}{k^{c_1}}\le \sup_{k\ge 1}\frac{\EE\left[\Din_v(S_v+k)+\deltain\right]}{k^{c_1}}<\infty.
\]

For $p\ge 2$, suppose 
$
\sup_{k\ge 1}\frac{\EE\left[(\Din_v(S_v+k)+\deltain)^r\right]}{k^{c_1 r}}\le A_r <\infty,
$
holds for some constants, $A_r$, $r=1,\ldots, p-1$.
Let $A_0 = \max\{A_r: r=1,\ldots, p-1\}$,
then by \eqref{eq:deg_p}, we have 
\begin{align}
\EE&\left((\Din_v(S_v+k+1)+\deltain)^p\right)\nonumber\\
&\le \EE\left((\Din_v(S_v+k)+\deltain)^p\left(1+p\frac{\alpha+\beta}{S_v+k+\deltain |V(S_v+k)|}\right)\right)\nonumber\\
&\qquad+ \sum_{r=1}^{p-1} (\alpha+\beta){p\choose 2}A_r k^{c_1 r-1}\nonumber\\
&\le \EE\left((\Din_v(S_v+k)+\deltain)^p\left(1+p\frac{\alpha+\beta}{S_v+k+\deltain |V(S_v+k)|}\right)\right)\nonumber\\
&\qquad + (\alpha+\beta)p(p-1)^2/2\, A_0 k^{c_1(p-1)-1}\nonumber\\
&=: C^{(1)}_v(k)+C^{(2)}_v(k).
\label{eq:induc}
\end{align}
We rewrite the $C^{(1)}_v(k)$ term in \eqref{eq:induc} to get
\begin{align*}
C^{(1)}_v(k) &= \left(1+\frac{c_1p}{k}\right) \EE\left[\left(\Din_v(S_v+k)+\deltain\right)^p\right]\\
&\quad +\EE\left[\left(\Din_v(S_v+k)+\deltain\right)^p\left(\frac{p(\alpha+\beta)}{S_v+k+\deltain |V(S_v+k)|}-\frac{c_1p}{k}\right)\right]\\
\le& \left(1+\frac{c_1p}{k}\right) \EE\left[\left(\Din_v(S_v+k)+\deltain\right)^p\right]\\
&\quad+\EE\left[\left(\Din_v(S_v+k)+\deltain\right)^p\frac{p\deltain||V(k)|-(1-\beta)k|}{(k+\deltain|V(k)|)(1+\deltain(1-\beta))k}\right]\\
\le&  \left(1+\frac{c_1p}{k}\right) \EE\left[\left(\Din_v(S_v+k)+\deltain\right)^p\right]\\
&\quad+\EE\left[\left(\Din_v(S_v+k)+\deltain\right)^p\frac{p\deltain||V(k)|-(1-\beta)k|}{k^2}\right].
\end{align*}
Similar to the Chernoff bound in \eqref{eq:chernoff}, we have for $p\ge 2$,
\begin{equation}\label{eq:chernoff2}
\PP\left(\Big|\big|V(k)\big|-(1-\beta)k\Big|\ge \sqrt{6p(1-\beta)k\log k}\right)
\le \frac{2}{k^{2p}}.
\end{equation}
Therefore, analogous to the calculation in \eqref{eq:deg1_bound}, we have
\begin{align*}
&\EE\left[\left(\Din_v(S_v+k)+\deltain\right)^p\frac{p\deltain||V(k)|-(1-\beta)k|}{k^2}\right]\\
&=\EE\left[\left(\Din_v(S_v+k)+\deltain\right)^p\frac{p\deltain||V(k)|-(1-\beta)k|}{k^2}\ind_{\left\{||V(k)|-(1-\beta)k|\le \sqrt{6p(1-\beta)k\log k}\right\}}\right]\\
&\, + \EE\left[\left(\Din_v(S_v+k)+\deltain\right)^p\frac{p\deltain||V(k)|-(1-\beta)k|}{k^2}\ind_{\left\{||V(k)|-(1-\beta)k|>\sqrt{6p(1-\beta)k\log k}\right\}}\right],\\
\intertext{and since $\left(\Din_v(S_v+k)+\deltain\right)^p\le (k+1+\deltain)^p$,}
&\le \EE\left[\left(\Din_v(S_v+k)+\deltain\right)^p\right]\frac{p\deltain \sqrt{6p\,k\log k}}{k^2}
+ \frac{p\deltain (k+1+\deltain)^p(k+1)}{k^2}\cdot\frac{2}{k^{2p}}.
\end{align*}
Hence,
\begin{align*}
C^{(1)}_v(k)&\le \EE\left[\left(\Din_v(S_v+k)+\deltain\right)^p\right]\left(1+\frac{c_1p}{k}+\frac{p\deltain \sqrt{6p\,k\log k}}{k^2}\right) \\
&\quad+
\frac{2p\deltain (k+1+\deltain)^p(k+1)}{k^{2p+2}}.
\end{align*}
Note also that $(k+1+\deltain)^p(k+1){k^{-2p-2}}\le k^{c_1(p-1)-1}$, for all $k\ge 1$, $p\ge 2$, and we conclude from \eqref{eq:induc} that
\begin{align*}
\EE&\left((\Din_v(S_v+k+1)+\deltain)^p\right)\\
&\le \EE\left((\Din_v(S_v+k)+\deltain)^p\right) \left(1+\frac{c_1 p}{k}+\frac{p\deltain\sqrt{6p k\log k}}{k^2}\right) \\
&\qquad+  \left(A_0(\alpha+\beta)p(p-1)^2/2+2p\deltain\right) k^{c_1(p-1)-1}.
\end{align*}
Following the recursive step as for the $p=1$ case gives
\[
\sup_{k\ge 1}\frac{\EE\left[(\Din_v(S_v+k))^p\right]}{k^{c_1 p}}\le \sup_{k\ge 1}\frac{\EE\left[(\Din_v(S_v+k)+\deltain)^p\right]}{k^{c_1 p}}<\infty.
\]

Note that for $p\ge 1$,
\begin{align*}
\EE\left(\left(\frac{\Din_v(n)}{n^{c_1}}\right)^p\right)
&= \EE\left(\left(\frac{\Din_v(n)}{n^{c_1}}\right)^p\ind_{\{S_v\le n-1\}}\right)
+\EE\left(\left(\frac{\Din_v(n)}{n^{c_1}}\right)^p\ind_{\{S_v\ge n\}}\right)\\
&\le \EE\left(\left(\frac{\Din_v(S_v+n)}{n^{c_1}}\right)^p\right)
+\left(\frac{1}{n^{c_1}}\right)^p,
\end{align*}
since $\Din_v(n)$ is monotone in $n$.
Then we have 
for $v\ge 1$,
\[
\sup_{n\ge 1}\EE\left(\left(\frac{\Din_v(n)}{n^{c_1}}\right)^p\right)<\infty.
\]
Applying a similar argument to the out-degrees completes the proof of the lemma.

\subsection{Proof of Lemma~\ref{lem:prelim}}\label{subsec:pf2.3}
We only show the results for $\Din_v(n)$, and those for $\Dout_v(n)$ follow from a similar argument.
First, by Lemma~\ref{lem:fil}, we see that for $n\ge 1$,
\begin{equation}\label{eq:mg_conv}
\frac{\Din_v(S_v+n)+\deltain}{\prod_{k=0}^{n-1}\left(1+\frac{\alpha+\beta}{S_v+k+1+\deltain|V(S_v+k)|}\right)}=:\frac{\Din_v(S_v+n)+\deltain}{X_n^{(v)}}
\end{equation}
is a non-negative $\left(\mathcal{F}_{S_v+n}\right)_{n\ge 0}$-martingale, which by the martingale convergence theorem, converges to some limit $L_v$ a.s. as $n\to\infty$. It remains to analyze the denominator and to verify $\PP(L_v=0)=0$. We do this by applying a similar proof machinery as in \cite[Lemma 8.17]{vanderHofstad:2017}.

By Markov's inequality, we see that for $\epsilon>0$, and $\max\{-1,-\deltain\}<m<0$,
\begin{align}
\PP(L_v\le \epsilon)&= \limsup_{n\to\infty}\PP\left(\frac{\Din_v(S_v+n)+\deltain}{X^{(v)}_n}\le \epsilon\right)\nonumber\\
&\le \epsilon^{|m|}\limsup_{n\to\infty} \EE\left[\left(\frac{\Din_v(S_v+n)+\deltain}{X^{(v)}_n}\right)^m\right]\nonumber\\
&\le \epsilon^{|m|}\limsup_{n\to\infty} \EE\left[\frac{\left(\Din_v(S_v+n)+\deltain\right)^m}{\prod_{k=0}^{n-1}\left(1+\frac{(\alpha+\beta)m}{S_v+k+1+\deltain|V(S_v+k)|}\right)}\right].
\label{eq:Lv}
\end{align}
By \eqref{eq:Lv}, it suffices to show
\[
\limsup_{n\to\infty} \EE\left[\frac{\left(\Din_v(S_v+n)+\deltain\right)^m}{\prod_{k=0}^{n-1}\left(1+\frac{(\alpha+\beta)m}{S_v+k+1+\deltain|V(S_v+k)|}\right)}\right]<\infty.
\]
Similar to \cite[Equation~(8.7.23)]{vanderHofstad:2017},  
there exists some constant $C_m$ such that
\begin{align*}
\limsup_{n\to\infty}\,& \EE\left[\frac{\left(\Din_v(S_v+n)+\deltain\right)^m}{\prod_{k=0}^{n-1}\left(1+\frac{(\alpha+\beta)m}{S_v+k+1+\deltain|V(S_v+k)|}\right)}\right]\\
&\le C_m \limsup_{n\to\infty}\,\EE\left[\frac{\Gamma(\Din_v(S_v+n)+\deltain+m)/\Gamma(\Din_v(S_v+n)+\deltain)}{\prod_{k=0}^{n-1}\left(1+\frac{(\alpha+\beta)m}{S_v+k+1+\deltain|V(S_v+k)|}\right)}\right].
\end{align*}
Hence, once we show 
\begin{equation}\label{eq:bound_gamma}
\limsup_{n\to\infty}\,\EE\left[\frac{\Gamma(\Din_v(S_v+n)+\deltain+m)/\Gamma(\Din_v(S_v+n)+\deltain)}{\prod_{k=0}^{n-1}\left(1+\frac{(\alpha+\beta)m}{S_v+k+1+\deltain|V(S_v+k)|}\right)}\right]<\infty,
\end{equation}
then the inequality in \eqref{eq:Lv} implies $\PP(L_v=0)=0$.

We prove \eqref{eq:bound_gamma} by showing
\[
M^{(m)}_n := \frac{\Gamma(\Din_v(S_v+n)+\deltain+m)/\Gamma(\Din_v(S_v+n)+\deltain)}{\prod_{k=0}^{n-1}\left(1+\frac{(\alpha+\beta)m}{S_v+k+1+\deltain|V(S_v+k)|}\right)},
\]
is a $\left(\mathcal{F}_{S_v+n}\right)_{n\ge 0}$-martingale.
Note that
\begin{align*}
&\EE^{\mathcal{F}_{S_v+n}}\left(\frac{\Gamma(\Din_v(S_v+n+1)+\deltain+m)}{\Gamma(\Din_v(S_v+n+1)+\deltain)}\right)\\
=& 
\frac{\Gamma(\Din_v(S_v+n)+\deltain+m)}{\Gamma(\Din_v(S_v+n)+\deltain)}\left(1-\frac{(\alpha+\beta)(\Din_v(S_v+n)+\deltain)}{S_v+n+1+\deltain|V(S_v+n)|}\right)\\
&+ \frac{(\alpha+\beta)(\Din_v(S_v+n)+\deltain)}{S_v+n+1+\deltain|V(S_v+n)|}
\frac{\Gamma(\Din_v(S_v+n)+\deltain+m)}{\Gamma(\Din_v(S_v+n)+\deltain)}\frac{\Din_v(S_v+n)+\deltain+m}{\Din_v(S_v+n)+\deltain}\\
=& \frac{\Gamma(\Din_v(S_v+n)+\deltain+m)}{\Gamma(\Din_v(S_v+n)+\deltain)}\left(1+\frac{(\alpha+\beta)m}{S_v+n+1+\deltain|V(S_v+n)|}\right),
\end{align*}
which confirms $M^{(m)}_n$
being a $\left(\mathcal{F}_{S_v+n}\right)_{n\ge 0}$-martingale.
Also,
\begin{align*}
\EE(M^{(m)}_n) &=
\EE(M^{(m)}_0)\\
&=
 \EE\left(\frac{\Gamma(\Din_v(S_v)+\deltain+m)/\Gamma(\Din_v(S_v)+\deltain)}{1+\frac{(\alpha+\beta)m}{S_v+1+\deltain v}}\right).
\end{align*}
Since $m<0$ and $S_v\ge v-1\ge 0$, then 
$\left(1+\frac{(\alpha+\beta)m}{S_v+1+\deltain v}\right)^{-1}\le \left(1+\frac{(\alpha+\beta)m}{(1+\deltain)v}\right)^{-1}$.
This further implies
\begin{align*}
\EE(M^{(m)}_n)
 &\le \left(1+\frac{(\alpha+\beta)m}{(1+\deltain)v}\right)^{-1}
 \EE\left({\Gamma(\Din_v(S_v)+\deltain+m)/\Gamma(\Din_v(S_v)+\deltain)}\right)\\
 &= \left(1+\frac{(\alpha+\beta)m}{(1+\deltain)v}\right)^{-1}
 \left(\alpha \frac{\Gamma(\deltain +m)}{\Gamma(\deltain)}
 + \gamma \frac{\Gamma(\deltain+1 +m)}{\Gamma(1+\deltain)}\right)<\infty,
\end{align*}
thus completing the proof of \eqref{eq:bound_gamma}.


Next, we consider the convergence of $X^{(v)}_n$
by noting that
\begin{align*}
\log& X^{(v)}_n\\
&= \sum_{k=0}^{n-1}\left[\log\left(1+\frac{\alpha+\beta}{S_v+k+1+\deltain|V(S_v+k)|}\right)
-\frac{\alpha+\beta}{S_v+k+1+\deltain|V(S_v+k)|}\right]\\
&\quad+\sum_{k=0}^{n-1} \left(\frac{\alpha+\beta}{S_v+k+1+\deltain|V(S_v+k)|}-\frac{c_1}{S_v+k+1}\right)\\
&\quad+\left(\sum_{k=0}^{n-1} \frac{c_1}{S_v+k+1}-c_1\log\frac{S_v+n}{S_v+1}\right)+c_1\log\frac{S_v+n}{S_v+1}\\
&=: I_v(n) + II_v(n) + III_v(n)+c_1\log\frac{S_v+n}{S_v+1}.
\end{align*}
Since $\log(1+x)\le x$, for all $x\ge 0$, then 
$I_v(n+1)-I_v(n)\le 0$ for all $n$, i.e.
$I_v(n)$ is decreasing in $n$. Note also that $|\log(1+x)-x|\le x^2/2$, for all $x\ge 0$, then we have
\begin{align*}
\EE&\left|\sum_{k=0}^{\infty}\left(\log\left(1+\frac{\alpha+\beta}{S_v+k+1+\deltain|V(S_v+k)|}\right)
-\frac{\alpha+\beta}{S_v+k+1+\deltain|V(S_v+k)|}\right)\right|\\
&\le \sum_{k=0}^{\infty}\EE\left|\log\left(1+\frac{\alpha+\beta}{S_v+k+1+\deltain|V(S_v+k)|}\right)
-\frac{\alpha+\beta}{S_v+k+1+\deltain|V(S_v+k)|}\right|\\
&\le \frac{(\alpha+\beta)^2}{2}\sum_{k=0}^{\infty}\EE\left(\frac{1}{S_v+k+1+\deltain|V(S_v+k)|}\right)^2\le \sum_{k=1}^{\infty}\frac{1}{k^2}<\infty,
\end{align*}
which implies $I_v(\infty)<\infty$ a.s., and $I_v(n)\convas I_v(\infty)$ as $n\to\infty$. 

By \cite[Theorem 3.9.4]{athreya:ney:2004}, we see that there exists a finite r.v. $Z$ such that
 \[
 \sum_{k=1}^\infty\left(\frac{\alpha+\beta}{k+\deltain |V(k-1)|}-\frac{c_1}{k}\right)\convas Z,
 \]
 then 
 $$II_v(n)\convas Z-\sum_{k=1}^{S_v}\left(\frac{\alpha+\beta}{k+\deltain |V(k-1)|}-\frac{c_1}{k}\right)
 =: II_v(\infty).$$
 Since $\sum_{k=1}^n 1/k-\log n\to \tilde{c}$, where $\tilde{c}$ is Euler's constant,
 then for $v=1$,
 $
 III_1(n) \convas c_1\tilde{c}=: III_1(\infty),
 $
 and for $v\ge 2$,
 \[
 III_v(n)\convas c_1\left(\tilde{c}+\log(S_v+1)-\sum_{k=1}^{S_v}\frac{1}{k}\right)=: III_v(\infty).
 \]
Hence, as $n\to\infty$,
\begin{equation}\label{eq:Xv_conv}
\frac{X^{(v)}_n}{\left((S_v+n)/(S_v+1)\right)^{c_1}}\convas \exp\left\{I_v(\infty)+II_v(\infty)+III_v(\infty)\right\}.
\end{equation}

Combining \eqref{eq:Xv_conv} with the convergence of the martingale in \eqref{eq:mg_conv}, we have 
 \[
 \frac{\Din_v(S_v+n)}{(S_v+n)^{c_1}}\convas \frac{L_v}{(S_v+1)^{c_1}}\exp\left\{-\bigl(I_v(\infty)+II_v(\infty)+III_v(\infty)\bigr)\right\}
=: \xi^\text{in}_v,
 \]
then
$$\lim_{n\to\infty}\frac{\Din_v(n)}{n^{c_1}}=\lim_{n\to\infty}\frac{\Din_v(S_v+n)}{(S_v+n)^{c_1}}
\stackrel{\text{a.s.}}{=} \xi^\text{in}_v.$$ 
Since $\PP(L_v=0)=0$ and $S_v+1\ge v\ge 1$, we also have $\PP(\xi^\text{in}_v=0)=0$.

\end{document}